\documentclass[aps,12pt,nofootinbib]{revtex4}
\usepackage{epsfig}
\usepackage{amsmath,amssymb}
\usepackage{longtable}
\usepackage{color}
\def\eq#1{Eq.~(\ref{#1})}
\def\fig#1{Fig.~\ref{#1}}
\def\sec#1{Sec.~\ref{#1}}
\def\tab#1{Tab.~\ref{#1}}

\begin{document}

\title[Coupling biochemistry and mechanics]{Coupling biochemistry and mechanics
in cell adhesion: a model for inhomogeneous stress fiber contraction}

\author{Achim Besser and Ulrich S. Schwarz}

\affiliation{University of Heidelberg, Bioquant, BQ 0013 BIOMS Schwarz,
Im Neuenheimer Feld 267, D-69120, Heidelberg, Germany}
\email{Ulrich.Schwarz@bioquant.uni-heidelberg.de}

\begin{abstract}
  Biochemistry and mechanics are closely coupled in cell adhesion.  At
  sites of cell-matrix adhesion, mechanical force triggers signaling
  through the Rho-pathway, which leads to structural reinforcement and
  increased contractility in the actin cytoskeleton. The resulting
  force acts back to the sites of adhesion, resulting in a
  positive feedback loop for mature adhesion. Here we
  model this biochemical-mechanical feedback loop for the special case
  when the actin cytoskeleton is organized in stress fibers, which are
  contractile bundles of actin filaments. Activation of myosin II
  molecular motors through the Rho-pathway is described by a system of
  reaction-diffusion equations, which are coupled into a viscoelastic
  model for a contractile actin bundle. We find strong spatial
  gradients in the activation of contractility and in the
  corresponding deformation pattern of the stress fiber, in good
  agreement with experimental findings.
\end{abstract}

\maketitle

\section{Introduction}

Adhesion of biological cells to each other and the extracellular
matrix is one of the hallmarks of multicellular organisms and a very
active area of research in cell biology. Investigating cell adhesion
is crucial to understand physiological processes like tissue
development and maintenance, but also disease-related processes like
growth and migration of cancer cells \cite{c:gumb96,c:huan99}.
In general, biological cells have a limited repertoire of possible
behaviours, including spreading, adhesion, migration, division or death,
but very sophisticated ways of controling the switching between these
different processes. Traditionally, investigation of this cellular
decision making has focused strongly on the biochemical aspects,
including detailed models for signal transduction
\cite{s:tyso03,s:khol06}. During recent years, it has become
increasingly clear that in adhesion-related processes, cellular behaviour is not
only controlled by biochemical cues, but also involves many physical
determinants like the structural organization of the extracellular
matrix and the cytoskeleton or force generation through molecular
motors \cite{c:disc05,c:voge06,uss:schw07a,c:lecu07}. For example, it
has been shown that the stiffness of the extracellular environment
determines migration of tissue cells \cite{c:lo00} and
differentation of stem cells \cite{c:engl06}. In particular, these celluar responses
have been found to depend on the ability of the cells to
contract their environment with actomyosin contractility
and to convert this mechanical process into
a biochemical signal.  Although these processes are
essential for such important situations like tissue functioning or
cancer cell migration, theoretical models describing the coupling
between biochemistry and mechanics in cell adhesion are still rare,
albeit essential for a future systematic understanding of how
multicellular organisms organize themselves.

Cell adhesion is closely related to the actin cytoskeleton, whose
organization is central in determining the structural properties of
cells. In cell culture with stiff substrates, the actin cytoskeleton
tends to organize in \textit{stress fibers}, which are bundles of actin
filaments tensed by myosin II molecular motors \cite{b:poll04}. Stress
fibers usually end in \textit{focal adhesions}, which are integrin-based
adhesion contacts which can grow to a lateral size of several microns
\cite{c:geig01a,c:bers03}. On their cytoplasmic side, focal adhesions
recruit more than 90 components (mostly proteins) which physically
reside in the adhesion structure \cite{c:zaid07}. In 1992, Ridley and
Hall published a landmark paper demonstrating that the assembly of
stress fibers and focal adhesions is regulated by a small GTPase
called Rho \cite{c:ridl92a}. Rho has many isoforms, but the one mainly
associated with focal adhesions is RhoA, which for simplicity in the
following we refered to as Rho. In a companion paper of the
same year, Ridley and Hall showed together with coworkers that another
small GTPase called Rac stimulates the formation of lamellipodia as
they appear in cell migration \cite{c:ridl92b}. The
main isoform associated with focal adhesions is Rac1 which for simplicity
in the following we will refer to as Rac.
While Rho mainly acts through activation of
actomyosin contractility, the main effect of Rac is activation of actin
polymerization, in particular activation of the actin nucleation
factor Arp2/3.  It has been reported later that activation of Rac
downregulates Rho, leading to disassembly of stress fibers and focal
adhesions \cite{c:sand99}. In many situations, Rho and Rac can be
regarded as antagonists, switching the cytoskeleton between different
structural states \cite{c:burr04}.  They are part of a larger family
of small GTPases, called the Rho-family, which for example also
includes Cdc42, which stimulates the formation of filopodia and
maintains cell polarity \cite{c:hall98}. Apart from regulation of the
actin cytoskeleton, the small GTPases from the Rho-family have many
other functions in the cell, for example in cell cycle control and
differentiation.

Although the small GTPases from the Rho-family are simple molecular switches,
they are regulated by many different factors. In general, GTPases are
upregulated by guanine nucleotide exchange factors (GEFs), which
convert the inactive Rho-GDP form to the active Rho-GTP form by
exchanging GDP for GTP. They are downregulated by GTPase-activating
proteins (GAPs), which stimulate Rho-GTPase activity, thus
leading to GTP-hydrolysis and transforming the active Rho-GTP to the
inactive Rho-GDP. For the 20 members of the Rho-family, 60 different
GEFs and 70 different GAPs as well as more than 60 different
downstream targets have been identified \cite{c:etie02}. At the
current stage of affairs, there is no way how this complex network can
be modeled in full detail. However, certain parts of this network have
been well characterized by biochemical assays, in particular
different parts of the Rho-mediated signal transduction pathway
leading from focal adhesions to actomyosin contractility.

Here we focus on the role of Rho as stabilizing factor for mature
adhesion. During recent years, it has been shown that Rho is the
central component of a biochemical-mechanical feedback loop which
regulates mature adhesion. In detail, it has been shown that
application of force on focal adhesions triggers their growth in a
Rho-dependent manner \cite{uss:rive01} (reviewed in
\cite{c:bers03,c:burr04}). Two main downstream targets of Rho leading
to stress fiber formation have been identified.  The formin mDia leads
to actin polymerization, while the Rho-associated kinase ROCK leads to
phosphorylation of myosin light chain and thus to increased motor
activity. Together these effects lead to formation of and
contractility in stress fibers and therefore to increased force levels
at focal adhesions. In this way, a positive feedback loop is closed
for upregulation of mature adhesion characterized by focal adhesions
and stress fibers. This biochemical-mechanical feedback loop is
schematically depicted in \fig{fig_overview}. An essential part of
this feedback loop is the growth of focal adhesion under force, which
recently has been the subject of different modelling approaches
\cite{Nicolas04,NicolasPRE04,Shemesh05,Besser06,Aroush06,Nicolas06}
(reviewed in \cite{Bershadsky07}). However, these models
have focused mainly on the mechanical and thermodynamic aspects of
the growth process, neglecting the interaction of mechanics and
biochemical signaling.  The positive feedback loop between
contractility and growth of adhesions has been modelled before in the
framework of kinetic equations, but without addressing the details of
force generation and its regulation by signaling pathways
\cite{Novak04}. Similar kinetic equations have been used to model the
antagonistic roles of Rho and Rac in cell adhesion, but again without
addressing the details of force generation and regulation
\cite{Scholey05}. Recently force generation has
been addressed in more detail in a model for whole cell contractility
and stress fiber formation \cite{c:desh06,c:desh07}. However, no
details of the signalling pathway have been modelled except for an
unspecified activation signal.

Because the actin cytoskeleton is very dynamic and interacts with many
different molecular factors, including actin-associated proteins and
molecular motors, it is very difficult to model its mechanical
properties in a general way. However, modelling becomes feasible if
one focuses on one of the well-characterized states of the actin
cytoskeleton, for example the lamellipodium or stress fibers.  Because
here we are mostly interested in mature cell adhesion in culture, we
will focus on the latter case. Modelling stress fibers can be
approached from different perspectives. An obvious starting point are
their common characteristics with muscle fibers, which is a linear
sequence of sarcomeres, each containing around 300 myosin II molecular
motors working collectively together as they slide the actin filaments
relatively to each other. This field has been pioneered by the
Huxley-model \cite{c:huxl57}, which later has been modified in many
regards, e.g.\ in regard to filament extensibility \cite{c:mija96} or
by a detailed modelling of the myosin II hydrolysis cycle
\cite{c:duke99}. In contrast to muscle fibers, stress fibers are more
disordered and a complete description therefore requires a model for
their assembly process from polar filaments interacting through
molecular motors. Such a description has been achieved in the
framework of a phenomenological theory which however does not model
the details of the underlying motor activity \cite{c:krus00,c:krus03}.
This theory does predict different dynamical states of the system,
including a stationary state of isometric contraction as observed in
stress fibers.

Although being less ordered than
muscle on the level of electron microscopy, stress fibers do exhibit a
periodic organization. \fig{fig_exp_data} shows
  experimental data for fibroblast adhesion to a stiff substrate
  \cite{c:pete04}. The image of a whole cell shown in
  \fig{fig_exp_data}A reveals a banding pattern in the stress fibers.
  The green regions correspond to the actin crosslinker
  $\alpha$-actinin while the red regions correspond to myosin II
  molecular motors.  Non-muscle myosin II is known to assemble into
  bipolar minifilaments consisting of 10-30 myosins, as depicted in
  the cartoon of a stress fiber in \fig{fig_overview}.  In contrast to
  muscle, the striation pattern of stress fibers shows considerable
  variability along the length of a stress fiber. In particular, it
  has been observed that upon stimulation of contraction with the drug
  calyculin A, only the sarcomeres in the periphery (close to the
  focal adhesions, \fig{fig_exp_data}B) shorten, while those in the
  center (close to the cell body, \fig{fig_exp_data}C) elongate. The
  exact time course of the striation width is shown in
  \fig{fig_exp_data}D and E for control and stimulation experiments,
  respectively.  This demonstrates that activation of contractility
  leads to strong spatial gradients in the striation pattern.  Thus
it is not sufficient to model only one sarcomeric unit as it is
typically done for muscle \cite{c:huxl57,c:mija96,c:duke99}. Rather at
least a one-dimensional chain of such sarcomeres has to be considered.

  Here we introduce a new one-dimensional model for
  stress fibers which essentially models them as linear sequence of
  viscoelastic Kelvin-Voigt bodies, whose stationary state is
  determined by the elastic part. The action of the
  molecular motors inside each sarcomeric unit are included on the
  level of a linearized force-velocity relationship. The signaling
  pathway is modeled as a system of reaction-diffusion equations.  For
  this study, we have conducted an extensive survey of the relevant
  literature and have collected the measured rate and diffusion
  constants in such a way that they now can be used for mathematical
  modelling.  The coupling between the biochemical signaling pathway
  and the mechanical stress fiber model proceeds by introducing a
  spatially varying fraction of active molecular motors. The local
  activation level is thereby determined by the outcome of the signaling
  pathway. A continuum limit of the mechanical model for many
  sarcomeric units in series results in a partial differential
  equation with mixed derivatives. The whole system of
  reaction-diffusion equations for the signal transduction and the
  partial differential equation for the mechanical part can be solved
  simultaneously. By feeding the force resulting from the mechanical
  model back into the activation of the signaling pathway, we obtain
  for the first time a model for the closed biochemical-mechanical
  feedback cycle described above. As a first application of our model,
  here we show that it predicts heterogenous contraction of stress
  fibers, in good agreement with experiments. In general, our work
  shows how models for the coupling of biochemistry and mechanics can
  be deviced in a meaningful way.

The paper is organized as follows. In \sec{sec:biochemistry} we start
with our model for the part of the Rho-pathway which is relevant for
our purposes. We then introduce our mechanical model for one
sarcomeric unit of the stress fibers in \sec{sec:mechanics} and its
continuum limit for a whole stress fiber in \sec{sec:continuum}. In
\sec{sec:feedback} our model predictions for inhomogeneous contraction
upon activation of contractility by calyculin are compared to the
experimental results. Finally we conclude in \sec{sec:conclusion}.

\section{Biochemical model for the Rho-pathway}
\label{sec:biochemistry}

In this study we concentrate on stress fibers and their regulation by
the Rho-pathway. In \fig{fig_overview} we introduce a
  coordinate system for our one-dimensional model: the stress fiber
  extends along the positive x-direction and the endpoints at $x=0$
  and $x=L$ correspond to two focal adhesions. Because the two focal
  adhesions are treated as equivalent, our model has inflection
  symmetry around $x=L/2$. In \fig{fig_biochemistry}, we schematically
  depict the biochemical part of our model in the spatial context of
  the focal adhesion at $x=0$ (by symmetry, the same description
  applies to the one at $x=L$).  Three compartments have to be
  considered: the focal adhesion, the cytoplasm and the stress fiber.
  In our model, each of these compartments corresponds to
  one or two important biochemical components. The reaction pathway is a
  linear sequence of activating or inihibitory enzyme reactions
  initiated at focal adhesion, transmitted through the cytoplasm by
  diffusion and resulting in spatially dependent myosin activation in
  the stress fiber. In the following we discuss each reaction step in
  detail and show how these processes are translated into
  reaction-diffusion equations. The abbrevations used for the
  biochemical components are compiled in \tab{tab_comp}, together with
  a short description of their functions. The model equations are
  summarized in \tab{tab_model_eq}.

  Our modelling of the biochemical signaling pathway
  starts with the activation of Rho at focal adhesions.  The
  Rho-protein has a lipophilic end that serves to anchor it to lipid
  membranes \cite{Seabra98}.  Complexation with guanine nucleotide
  dissociation inhibitors (GDIs) shields the hydrophobic parts of the
  Rho-protein and make it inactive as well as soluble in the cytoplasm
  \cite{Somlyo00}. It is expected that Rho is released from these
  complexes at focal adhesions.  More importantly, focal adhesions are
  known to recruit different Rho-GEFs, thus activating Rho at the
  focal adhesions. The active Rho is then able to bind Rho-associated
  kinase (ROCK) and thereby activate its kinase activity
  \cite{Feng99}. Since the active ROCK is bound to Rho-GTP, we assume
  in our model that these components are not diffusible but are
  localized to the focal adhesions. For the same reason, we neglect
  the direct interaction of ROCK and myosin which has been reported to
  occur \textit{in vitro}.  Active ROCK phosphorylates the diffusible
  myosin light chain phosphatase (MLCP) at its myosin-binding subunit
  (MBS). MLCP and its antagonistic partner myosin light chain kinase
  (MLCK) are the main regulators of myosin contractility in the
  context we are interested in. Both enzymes interact with the
  regulatory myosin light chain subunit (MLC) of the myosins.
  Depending on the respective activities, MLC gets either
  phosphorylated or dephosphorylated.  MLC, in turn, controls the
  myosin binding to actin filaments. Only if MLC is phosphorylated
  myosin is able to bind actin filaments and perform its ATPase cycle
  that converts chemical energy into mechanical work, e.g.
  contraction of stress fibers \cite{Kamm85}. By phosphorylating MLCP,
  ROCK effectively enhances the phosphorylation level of MLC. In this
  way, Rho-activation can lead to an increase in myosin contractility
  in the stress fiber.

Our model for the biochemical reaction-diffusion system assumes that
each enzyme stimulation follows Michaelis-Menten kinetics
\cite{Murray02}. In Michaelis-Menten kinetics,
production first increases linearly with educt concentration
and then saturates at a maximal production velocity if educt
concentration exceeds the value set by the Michaelis-Menten constant.
The precise molecular process corresponding to the
conversion of force into a biochemical signal at focal adhesions has not been
identified yet.
However, it is expected that mechanical forces exerted onto focal
adhesions eventually initiate the loading of Rho-GTP leading to ROCK
activation. For lack of information we therefore lump the focal
adhesion associated processes into one equation that effectively
describes the conversion of ROCK into its activated form (presumably
complexed with Rho-GTP). The mechanical force $F_{b}$ that stimulates
the activation is treated as enzyme in the framework of
Michaelis-Menten kinetics:
\begin{equation}
\label{eq_ROCK}
\frac{\partial ROCK(t)}{\partial t}=\frac{r_1
  F_{b}(t)(ROCK_{tot}-ROCK(t))}{K_1+ (ROCK_{tot}-ROCK(t))} -\frac{V_{-1}
  ROCK(t)}{K_{-1}+ROCK(t)}
\end{equation}
  The variable $ROCK$ denotes the activated form of ROCK
  and we assume that the overall concentration of ROCK is constant at
  $ROCK_{tot}$.  The force exerted by the stress fiber onto the focal
  adhesion, $F_{b}(t)$, stimulates the conversion of ROCK into its
  activated form with maximum velocity $r_1 F_{b}(t)$ and
  Michaelis-Menten constant $K_{1}$. The parameter $r_1$ is equivalent
  to a rate constant but relates mechanical force to a chemical
  reaction. For this reason the units of $r_1$ are given as
  $[nM/s\,nN]$. The force $F_{b}(t)$ will depend on the
  stress fiber deformation. The second term accounts for the
degradation of activated ROCK to its inactive form,
with maximum velocity $V_{-1}$ and Michaelis-Menten
  constant $K_{-1}$.  Since we expect ROCK in its active form to be
associated with focal adhesions, we omit diffusive contributions to
this equation.

One main effector of ROCK is MLCP which we regard as a diffusible compound leading to a
reaction-diffusion equation:
\begin{equation}
\label{eq_Pt}
\frac{\partial \textit{MLCP}(x,t)}{\partial t}=\frac{V_{-2}
  \textit{MLCP-P}(x,t)}{K_{-2}+\textit{MLCP-P}(x,t)}+D\frac{\partial^2
\textit{MLCP}(x,t)}{\partial x^2}
\end{equation}
  Here, the variables $\textit{MLCP-P}$ and $MLCP$
  denote the phosphorylated and unphosphorylated form of myosin light
  chain phosphatase, respectively.  The first term accounts for the
  dephosphorylation of MLCP-P with maximum velocity $V_{-2}$ and
  Michaelis-Menten constant $K_{-2}$. The second term allows for the
  diffusion of the phosphatase with diffusion constant $D$. The
phosphorylation level of MLCP is also regulated by the active form of
ROCK which catalyzes the reverse reaction, that is the conversion of
the phosphatase into its phosphorylated form. However, ROCK is only
active in the vicinity of focal adhesions located at each end of the
stress fiber. Therefore this source term can be incorporated into the
boundary conditions for \eq{eq_Pt}, in the sense that the diffusive
flux into the boundary has to balance the conversion into its inactive
form:
\begin{equation}
\label{eq_BC_Pt}
D \frac{\partial \textit{MLCP}(x=0,t)}{\partial x}= \frac{R_2 ROCK(t)
\textit{MLCP}(x=0,t)}{K_2+\textit{MLCP}(x=0,t)}
\end{equation}
The same relation, but with inverted sign is valid at the other end at
$x=L$, compare the overview in \tab{tab_model_eq}.
The reaction is again modelled with Michaelis-Menten kinetics, where
$R_2=r_2 v_{b}$ is the product of a rate constant $r_2$ with an
effective volume $v_{b}$ of the focal adhesion in which the reaction
takes place.  $K_2$ is the usual Michaelis-Menten constant. For the
phosphorylated form of the phosphatase (MLCP-P) equations similar to
\eq{eq_Pt} and \eq{eq_BC_Pt} are valid, but with inverted signs of the
source terms and a diffusion constant which in principle can be
different, see \tab{tab_model_eq}.

MLCP together with MLCK regulate the phosphorylation level of MLC. Since myosin in
stress fibers form mini-filaments which are bound to actin filaments, we
neglect diffusion of this compound, leading to the rate equation for the
phosphorylated fraction $n$ of MLC:
\begin{equation}
\label{eq_n}
\frac{\partial n(x,t)}{\partial t}=\frac{V_3
  \left(1-n(x,t)\right)}{K_3+(1-n(x,t))} -\frac{r_{-3} \textit{MLCP}(x,t)
  n(x,t)}{K_{-3}\,\,I+n(x,t)}
\end{equation}
By allowing only the ratio of the phosphorylated fraction
to vary, we assume that the overall amount of myosin in the stress fibers is fixed.
MLC is phosphorylated by MLCK with a maximum velocity $V_3=r_3 MLCK$
and respective Michaelis-Menten constant $K_{3}$. Here we assume that
the concentration of MLCK is constant within the cell. The
kinase is antagonized by MLCP that dephosphorylates MLC with a rate
constant $r_{-3}$ and Michaelis-Menten constant 
$K_{-3}$. The factor $I$ is an inhibition parameter defined below. Since MLCP
has spatial dependent source terms and is diffusible, the inhibition
of MLC by the phosphatase will vary in space.

To complete the biochemical modelling we have to specify how the
induction of calyculin is treated in our model. Calyculin is an
inhibitor of MLCP and thereby enhances the phosphorylation level of
MLC. We model the interaction of calyculin with its target MLCP as a
competitive inhibition leading to the additional factor $I$ in the
last term of \eq{eq_n} \cite{SysBioInPrac05}. In the presence of
calyculin $I>1$ (in absence of the drug: $I=1$) which effectively
increases the Michaelis-Menten constant $K_{-3}$ and thus decreases
the rate of MLC dephosphorylation. Hence more myosin motors will be
activated and cell contractility is stimulated. The induction of the
drug is then modelled by switching instantaneously the inhibition
parameter from $I=1$ to $I=3$. Thereby we omit the time delay
caused by the internalization of the drug. 

The used parameter values for the reaction-diffusion system are based
on an extensive survey of the literature and are summarized in
\tab{tab_parameters}. If a range of values is reported
  in the literature, we chose an intermediate value for this parameter. If no
  value could be found in the literature, we made reasonable
  assumptions based on similar parameters in other systems. No attempt
  was made to fit the parameters to some target function. We first
analyze the properties of this reaction diffusion system assuming that
the boundary force exerted onto the focal adhesions is held at a
constant level. This would be the case if the myosin forces were in a
stationary state and not regulated by the biochemical signals emerging
from focal adhesions. We impose artificial initial conditions that all
components ($ROCK$, $\textit{MLCP-P}$ and $n$) are at zero activation
level but set the boundary force to 5 nN, a typical force observed for
fibroblast \cite{uss:bala01}. This mechanical stimulation triggers the
accumulation of active ROCK at focal adhesions, see \eq{eq_ROCK},
creating a sink for the active form of MLCP. Thus in those boundary
regions MLCK dominates and increases MLC phosphorylation. Closer to
the center of the cell MLC rather remains in its unphosphorylated
form. The width of the interfacial region of intermediate
MLC-phosphorylation level is mainly determined by the diffusiveness of
the phosphatase. The faster the diffusion, the wider the intermediate
region. On a typical timescale of a few minutes all components
equilibrate to their steady state concentration profile, where MLC is
highly phosphorylated at the boundaries, but is poorly activated at
the center of the stress fiber. In \fig{fig_n_para_plot} we show the
typical equilibration of the phosphorylated fraction of MLC, $n(x,t)$,
as obtained from a solution of the full system of biochemical
reaction-diffusion equations. Below we will argue
  that this phosphorylation profile of MLC implies a spatially varying
  myosin motor activation leading to an inhomogeneous stress fiber
  contraction.

\section{Sarcomeric unit of stress fiber model}
\label{sec:mechanics}

A minimal model for stress fibers has to take into account not only
the viscoelastic but also the contractile properties of the fiber due
to myosin motor activity. For the mechanical response of a sarcomeric
unit we take the usual Kelvin-Voigt model for viscoelastic material
\cite{b:fung93}. It consists of a dashpot with viscosity $\eta$ and a
spring of stiffness $k$ connected in parallel. These two modules
represent viscous and elastic properties of the material,
respectively. The Kelvin-Voigt model is the simplest
  viscoelastic model which in the stationary state is determined by
  elasticity, in contrast to the Maxwell model, which flows in the
  stationary limit. Thus the Kelvin-Voigt model is the appropriate choice for
  stress fibers, which can carry load at constant deformation over a long
  time. In order to cope with the contractile behavior of stress
fibers, we introduce an additional contractile element that represents
the activity of motor proteins. For illustrative reasons we first
derive the governing differential equation for such a viscoelastic and
contractile element (\textit{vec-element}) before we proceed to the
stress fiber model. This ansatz is similar to the two-spring model
which we have introduced before to explain the physical aspects of
rigidity sensing on soft elastic substrates \cite{uss:schw06a}.

The considered vec-element is depicted in \fig{fig_vec-element}.
The properties of the additional contractile element is given by
the specific force-velocity relation of the molecular motor. For
simplicity, we use the linearized relationship:
\begin{equation}
\label{eq_FVR}
F_m(v)=F_{stall}\left(1-\frac{v}{v_0}\right)
\end{equation}
$F_m$ is the actual force exerted by a motor moving with velocity
$v$. $v_0$ is the zero-load velocity and $F_{stall}$ is the stall
force of the motor, that is the maximal force
allowing motor movement. In the following description for the stress
fiber, the stall force will depend on the active fraction of the
myosin, compare \eq{eq_Fstall}, such that myosin contractility may
vary spatially. The sum of all internal forces, F, exerted by the vec-element reads
\begin{equation}
\label{eq_force_balance_vec}
F=-\eta \frac{du}{dt}-ku-F_m
\end{equation}
The crucial point is that the motor force $F_m(v)$ is related to the
contraction velocity $v$ which relates to the displacement $u(t)$
simply by
\begin{equation}
\label{eq_v}
v=-\frac{du}{dt}
\end{equation}
The minus sign comes from the fact that the directed motor
movement causes a relative sliding of the anti-parallel orientated
actin filaments leading to contraction of the element with
velocity $v$. Thus the displacement $u(t)$ becomes negative upon
motor activity. If there are no external forces, all internal forces have to balance, $F=0$, and 
we arrive at the following differential equation for the displacement u(t):
\begin{equation}
\label{eq_ode_vec}
(\eta+\frac{F_{stall}}{v_0}) \frac{du}{dt}+ku=-F_{stall}
\end{equation}
The term originating form the motor activity decomposes into two
contributions. First it increases the viscosity of the damping
term (dissipative signature of the motor) and second it
contributes a constant force to the inhomogeneous part of the
equation (contractility of the motor). Assume vanishing initial
displacement $u(t=0)=0$, the solution to equation \eq{eq_ode_vec}
is given by:
\begin{equation}
\label{eq_sol_vec}
u(t)=-\frac{F_{stall}}{k}\left(1-e^{-t/\tau}\right)
\end{equation}
Here $\tau=\eta_{e}/k$ corresponds to the relaxation time of the
Kelvin-Voigt model with an effective viscosity
$\eta_{e}=\eta+F_{stall}/v_0$. Thus the system appears to be more
viscous for stronger and slower motors and subsequently the motor
activity slows down any relaxation due to perturbations of the system.
The steady state deformation $u(t\to\infty)=-F_{stall}/k$ is
determined by the ratio of stall force and spring stiffness.

\section{Continuum version of stress fiber model}
\label{sec:continuum}

We model a stress fiber as a string of vec-elements whose
viscoelastic and contractile properties may vary spatially. For
example the contractile strength of the vec-elements will vary
spatially because the biochemical signal will cause different
myosin activation levels along the fiber. Starting from a discrete
description depicted in \fig{fig_springs_dashpots_and_motors} we
derive the governing continuum equation. Similar to the preceding
discussion, the force $F_n$ on the site $n$ is the sum of spring
forces, viscous drag and the forces built up by the motor
proteins:
\begin{equation}
\label{eq_Fn}
F_n=\eta_{n+1}\frac{\partial}{\partial
  t}(u_{n+1}-u_n)-\eta_n\frac{\partial}{\partial
  t}(u_n-u_{n-1})+k_{n+1}(u_{n+1}-u_n)-k_n(u_n-u_{n-1})+F_{m_{n+1}}-F_{m_{n}}
\end{equation}
Here, we allow that the stiffness of the spring, the viscosity as
well as the motor force vary spatially. In order to deduce a
continuum description we expand the functions $\eta,k,u$ and $F_m$
at $x=na$ assuming that these functions are smooth within the
small distance $a$.  Keeping only leading order terms yields:
\begin{equation}
\label{eq_F}
F(x,t)=a^2\left(\frac{\partial}{\partial x} \eta(x)
  \frac{\partial}{\partial x} \frac{\partial}{\partial
    t}+\frac{\partial}{\partial x} k(x)\frac{\partial}{\partial
    x}\right)u(x,t)+a\frac{\partial F_m(x,t)}{\partial x}
\end{equation}
Note that the leading differential operator
$\partial_x$ acts on $\eta$ and $u$. The same holds
for the second term. If $k$ does not vary spatially, it
simplifies to $k \partial_x^2$. Like
for a single vec-element, we argue that the motor force $F_m$ depends
on the displacement $u(x,t)$. The contraction $\Delta_n$ within the
n-th element generated by the respective motor is given by $\Delta_n =
- (u_n-u_{n-1})$. The contraction velocity is therefore
$v(x,t)=\dot{\Delta}(x,t) \approx -a\partial_t \partial_x u(x,t)$.  The
found expression for the velocity is inserted into the force velocity
relation \eq{eq_FVR} leading to:
\begin{equation}
\label{eq_FVR_xt}
F_m(x,t)=F_{stall}(x,t) \left( 1+ \frac{a}{v_0} \frac{\partial}{\partial t}
  \frac{\partial}{\partial x} u(x,t) \right)
\end{equation}
In contrast to \eq{eq_FVR}, here the stall force is not constant but depends on
the phosphorylated fraction $n(x,t)$ of MLC along the stress fiber. 
This comes from the fact that along a
myosin minifilament and depending on MLC phosphorylation, a larger
or smaller fraction of myosin heads is able to bind to actin and
perform ATP-cycles. The more myosin heads are active the larger
the maximum force that the bundle can exert to the actin
filaments. In our model we regard the ensemble of myosins within a
cross-section of a stress fiber as one large contractile unit with
an effective stall force that depends linearly on the active
fraction $n$ of myosin heads:
\begin{equation}
\label{eq_Fstall}
    F_{stall}(x,t)=F_{max}\,n(x,t)
\end{equation}
The effective stall force, $F_{stall}(x,t)$, would reach the
maximum force $F_{max}$ if all myosins within this cross section
would be working ($n=1$). In the following we set $F_{max}=50$ nN
which will result in boundary forces exerted by the fiber of about
$5$ nN which corresponds to typical values observed in experiments
\cite{uss:bala01}.
\eq{eq_F} together with \eq{eq_FVR_xt} and \eq{eq_Fstall} lead to 
the final model equation for the stress fiber:
\begin{equation}
\label{eq_boxed}
\left(\frac{\partial}{\partial
      x}\eta_{e}(x,t)\frac{\partial}{\partial x}
    \frac{\partial}{\partial t}+\frac{\partial}{\partial x} k(x)
    \frac{\partial}{\partial
      x}\right)u(x,t)=-\frac{1}{a}\frac{\partial}{\partial x}
  F_{stall}(x,t)
\end{equation}
where the effective viscosity
$\eta_{e}(x,t)=\eta(x)+F_{stall}(x,t)/v_0$, similar to the
findings in \eq{eq_ode_vec}, although here the viscosity varies
spatially. Interestingly, only the variation of the motor force
appears on the right hand side of the equation. As a consequence,
a homogeneous motor activity will not contribute to the
displacements within the string.

To obtain some intuition for this equation, assume that the
spatially varying stall force is given e.g. by the steady state
solution $n_{ss}$ of the reaction diffusion system, depicted in
\fig{fig_n_para_plot}, such that $F_{stall}(x)=F_{max}n_{ss}(x)$.
For this simplifying case where the stall force does not vary in
time, \eq{eq_boxed} can be integrated and the time dependent
solution for $u(x,t)$ is given by:
\begin{align}
    u(x,t) = & \,\,u(x_0,t)+\int^x_{x_0}dx'\left(
\partial_{x'}u(x',t_0)e^{-\frac{t-t_0}{\tau(x')}}-\frac{F_{stall}(x')}{a
    k(x')}
\left[1-e^{-\frac{t-t_0}{\tau(x')}}\right] \right. \nonumber \\
\label{eq_ana_sol}
    & \left. +
\frac{1}{a\eta_{e}(x')}e^{-\frac{t-t_0}{\tau(x')}}\int^t_{t_0}F_{b}(t')e^{\frac{
t'-t_0}{\tau(x')}}dt'\right)
\end{align}
Here, we have set $\tau(x)=\eta_{e}(x) / k(x)$, the typical
relaxation time with which perturbations decay at a certain
position. E.g. the initial conditions $\partial_{x}u(x,t_0)$ can be regarded
as perturbations to the steady state and they decay with
$exp(-t/\tau(x))$. The three integration constants can be
identified as the displacement at the left boundary, $u(x_0,t)$,
the force exerted to the left boundary, $F_{b}(t)$, and the
initial strain along the fiber, $\partial_{x}u(x,t_0)$. They are determined
by the boundary and initial conditions. Experiments by 
Peterson et al.\ \cite{c:pete04}
are arranged with cells on stiff substrates to which the ends of
the stress fiber are connected by focal adhesions. Therefore, the
appropriate boundary conditions are fixed ends for the fiber,
namely $u(x_0,t)\equiv0$ and $u(x_e,t)\equiv0$. From the second
condition one is able to calculate the missing integration
constant $F_{b}(t)$ for any initial condition $\partial_{x}u(x,t_0)$. The
force on the left boundary $F_{b}(t)$ is given as solution to an
inhomogeneous Volterra equation of the first
  kind:
\begin{equation}\label{eq_volterra}
\int^t_{t_0}K(t-t')F_{b}(t')dt'=g(t)
\end{equation}
with the kernel
\begin{equation}
\label{eq_kernel}
K(t-t')=\int^{x_e}_{x_0}dx'\frac{1}{a\eta_{e}(x')}e^{-\frac{t-t'}{\tau(x')}}
\end{equation}
and inhomogeneous part $g(t)$ dependent on the initial condition $\partial_{x}u(x,t_0)$:
\begin{equation}
\label{eq_inhom}
g(t)=\int^{x_e}_{x_0}dx'\left(\frac{F_{stall}(x')}{a
    k(x')}\left[1-e^{-\frac{t-t_0}{\tau(x')}}\right]-
  \partial_{x'}u(x',t_0)e^{-\frac{t-t_0}{\tau(x')}}\right)
\end{equation}
In order to solve the integral equation \eq{eq_volterra} for
$F_{b}(t)$, we calculate its time derivative, leading to:
\begin{equation}
\label{eq_dt_volterra}
F_{b}(t)=\frac{\dot{g}(t)}{K(0)}-\frac{1}{K(0)}\int^t_{t_0}\dot{K}(t-t')F_{b}
(t')dt'
\end{equation}
This equation yields an explicit expression for the initial
force $F_{b}$ at t=0:
\begin{equation}
\label{eq_Flb0}
F_{b}(t_0)=\frac{\dot{g}(t_0)}{K(0)}\neq0
\end{equation}
By inspection of the kernel \eq{eq_kernel} and the inhomogeneous
part \eq{eq_inhom} one finds that the initial force onto the
boundary has a finite value, even for the initial condition
$\partial_{x}u(x,t_0)=0$. \eq{eq_dt_volterra} also yields an iteration rule
for the time course of $F_{b}(t)$, by applying a quadrature,
where $F_{b}(t_0)$ from \eq{eq_Flb0} is used as a starting value.
The solution for $F_{b}(t)$ is shown in \fig{fig_fb_ana_num}. The
boundary force is rising from its initial value and then quickly
saturates at about $4.8$ nN. The result for $F_{b}(t)$ can then be
set into the general solution (\eq{eq_ana_sol}) for the
displacement $u(x,t)$ along the fiber. \fig{fig_u_para_plot} shows
this solution, by using the steady state activation level for the
myosins ($F_{stall}(x)=F_{max}n_{ss}(x)$) shown in
\fig{fig_n_para_plot} and assuming the initial condition
$\partial_{x}u(x,t_0)=0$ as well as the boundary conditions $u(0,t)\equiv0$
and $u(L,t)\equiv0$. Beside the analytical solution, indicated by
circles, we also included the direct numerical solution of
\eq{eq_boxed} for comparison. The numerical solution was derived
by using the MATLAB algorithm "pdepe". The sinusoidal shape of the
function $u$ results from stronger contractile motors close to the
boundaries causing the fiber elements to displace into the
direction of the boundaries. Hence the displacement $u$ is
positive (negative) along the right (left) half of the fiber. It
is worth noting that the mechanical equilibration of the stress
fiber occurs within seconds in contrast to the biochemical
system which equilibrates over minutes.

\section{Role of feedback}
\label{sec:feedback}

We already argued that the system of focal adhesions and stress fibers
exhibit a closed biochemical and mechanical positive feedback loop.
Despite this fact the previous results were derived under the
assumption that the mechanically triggered biochemical signals at FAs
originate from a constant force. In order to model the full biological
system, the varying boundary forces have to be fed back into the
equation describing the mechanotransduction \eq{eq_ROCK}. Since the
stress fiber model does not include any cross-links (e.g. intermediate
contacts to the substrate) the tension $\gamma$ within the fiber has
to be constant and therefore equals the boundary forces:
\begin{equation}
\label{eq_line_tension}
F_{b}(t)=\gamma(x=0)= a\eta_{e}(0,t)\partial_x \partial_t u(0,t)+ak(0)
\partial_x u(0,t)+F_{stall}(0,t)
\end{equation}
This relation now connects the biochemical signaling
to the mechanical deformation of the stress fiber.
Thus, the coupled system of reaction equations, \eq{eq_ROCK} to
\eq{eq_n}, and the mechanical equation \eq{eq_boxed} have to
be solved simultaneously. This can be done numerically by using
the MATLAB algorithm "pdepe". The whole system of equations 
and the used parameter values are summarized in \tab{tab_model_eq}
and \tab{tab_parameters}, respectively.

By doing a steady state analysis we find, that this system of
equations exhibit two stable steady states for the used parameter
values: the first state is characterized by a generally low activation
level $n_{ss}(x)$ of the myosin motors resulting in marginal boundary
forces whereas in the second state myosin motors are non-uniformly
activated and the exerted forces reach a few $nN$. This bistability is
characteristic for a positive feedback system \cite{s:tyso03}. The
first "non-active" state would correspond to cells that failed to
establish mechanical stress whereas the second "active" state
correspond to cells that are well adhered to the substrate. In order
to simulate the drug experiments by Peterson et al.\
  \cite{c:pete04}, we start with the system residing in this "active"
state and then, at t=0, we perturb the system by turning on the
stimulation with calyculin. This is modelled by switching
instantaneously the inhibition parameter from $I=1$ to $I=3$, thereby
omitting the time delay caused by the internalization of the drug. The
stimulation with calyculin reduces the phosphatase activity and
elevates the myosin activation level everywhere leading to a quick
increase in the boundary forces exerted by the stress fiber. The time
course of the force exerted onto the focal adhesions is shown in
\fig{fig_fb_inhib}. Subsequently, the positive mechanical feedback
triggers additional signaling at focal adhesions activating myosin
motors preferentially at the cell periphery. This results in strong
spatial gradients in myosin motor activity, see \fig{fig_nss_inc_wo}.
The strong peripheral motors then contract the fiber to the cost of
the central regions where the fiber has to elongate. This can be
further analyzed by using the numerical solution for the displacement
$u(x,t)$. The steady states of the displacement $u$ before and after
stimulation with calyculin are shown in \fig{fig_u_b_n_a}. The
stimulation strongly increases the displacement along the fiber
resulting in substantial contraction of the fiber close to the
boundaries but in expansion around the cell center. This finding
becomes more apparent in the shown striation pattern calculated from
the displacement after stimulation (upper string) compared to the
striation pattern of a completely undistorted fiber (lower string).
The bands close to the boundaries have been contracted whereas the
bands around the center have been expanded, compare also
\fig{fig_mean_bw}.  We have to stress that the presented stress
fiber model is continuous, thus the model cannot distinguish between
$\alpha$-actinin bands or MLC bands. The color code in
\fig{fig_u_b_n_a} and \fig{fig_mean_bw} is therefore arbitrary. We also derive the
local relative change of density within the fiber which is given in
general as the negative trace of the strain tensor
$\delta\rho_{rel}=(\rho-\rho_0)/\rho_0=-tr(u_{i,j})$. Since the model
is one dimensional this simplifies to
$\delta\rho_{rel}=-\partial_{x}u(x,t)$, plotted in
\fig{fig_density_inc_wo}.  The local relative change of band width at
a certain position within the fiber, is then simply given by:
$(w(x,t)-w_0)/w_{0}=-\delta\rho_{rel}(x,t)=\partial_{x}u(x,t)$.
The figure shows that the inhomogeneous motor activity causes a
contraction of the bands up to about 55\% close to the fiber ends (the
relative change in density is positive), whereas the pattern expand up
to 15\% at the middle of the fiber (the relative change in density is
negative).

The experimental time course data for the sarcomer length shown in
\fig{fig_exp_data} is intrinsically averaged over a certain area in
the peripheral and central regions of the cell. In order to compare
the model results with the experimental finding we therefore define
central (center $\pm10\mu m$) and peripheral (edges $\pm10\mu m$)
regions of the cell, indicated by vertical lines in
\fig{fig_density_inc_wo}. The expected sarcomer length at a certain
position along the fiber is given by
\begin{equation}
w(x,t)=w_0+u(x+w_0,t)-u(x,t) \,\,\, or
\,\,\, w(x,t)=w_{0}\left(1+\partial_{x}u(x,t)\right) \,\,\,for \,\,\, w_0\ll L
\end{equation}
In the following analysis this measure is averaged over the defined
central and peripheral regions, respectively. The deduced time courses
for the mean pattern bandwidths in the distinct regions are shown in
\fig{fig_mean_bw}. The expected steady state striation patterns are
illustrated as insets. Upon stimulation with calyculin, the peripheral
mean bandwidth shrinks from its initial value of about $0.97\ \mu m$
down to $0.83\ \mu m$, whereas in the central regions, the bands
elongate from about $1.03\ \mu m$ up to $1.13\ \mu m$. Interestingly
the inital mean bandwidth at the center and periphery yet differ in
the initial unperturbed steady state of the cell ($1.03\ \mu m$
compared to $0.97\ \mu m$). This results from the fact that the
unperturbed fiber already exerts moderate forces on to the focal
adhesions which results in slight spatial gradients in myosin
activation. These gradients then sharpen upon stimulation with
calyculin, see \fig{fig_nss_inc_wo} as well as
\fig{fig_density_inc_wo}. The model results agree qualitatively with
the experimental findings by Peterson et al.\
  \cite{c:pete04} and the quantitative measurements are within the
same order of magnitude (compare \fig{fig_exp_data}). It is worth
mentioning that the amplitude of contraction or elongation of the
fiber scales inversely with the fiber stiffness $k$: the softer the
fiber, the stronger the mechanical deformation will be. Thus, a lower
$k$ value would simply explain the reported higher values for sarcomer
contraction of about 30-40\%. In our calculation we used $k=45nN/\mu
m$, a value reported by \cite{Deguchi06}. The experimentally measured
equilibration time of the stress fiber upon stimulation is about 20
min (\fig{fig_exp_data}) which compares to about 3 min for the model
results. These quantitave difference originate from
two model simplifications. First, we lump the focal adhesion
associated processes into one equation thereby we shortcut the
activation of ROCK and neglect prior activation steps of e.g. Rho-Gef
or Rho-GTPase.  Considering these steps would cause an additional time
delay.  Secondly, the stimulation with calyculin happens
instantaneously in the model omitting the time delay caused by the
internalization of the drug. Refining the model and eliminating these
simplifications will further decrease the differences in equilibration
times.

To highlight the importance of the mechanical feedback we also include
the expected results for a system neglecting this feedback shown as
dashed lines in \fig{fig_nss_inc_wo} and \fig{fig_density_inc_wo}.
Here the homogeneous induction of the drug cause an almost uniform
elevation of myosin activation within the cell. Slight differences
between cell center and cell periphery would persist but rather
marginally extend (\fig{fig_nss_inc_wo}). In fact stimulation leads
also here to amplified distortions of the striation pattern. However,
the changes in bandwidths are significantly smaller compared to the
system incorporating the feedback (\fig{fig_density_inc_wo}).
Thus the closed biochemical and mechanical feedback
  loop is an essential feature required to describe the strong
  distortions of striation patterns upon homogenous drug induction.

\section{Conclusion}
\label{sec:conclusion}

Here we have presented for the first time a mathematical model for the
closed biochemical-mechanical feedback loop triggering the
upregulation of focal adhesions and stress fibers which is typical for
cell culture on stiff substrates. In regard to the biochemical part,
we present for the first time a reaction-diffusion model for
Rho-signaling from focal adhesions towards stress fibers.  Our
modelling is based on an extensive review of the literature, which
provides the list of diffusion and reaction constants summarized in
\tab{tab_parameters}. In regard to the mechanical part, we introduced
a new model for stress fibers which takes into account the special
viscoelastic and contractile properties of the sarcomeric units. For a
linear chain of many such units, we derived a continuum equation which
we solved both analytically and numerically. Combining the two model
parts resulted in a complete model for the feedback loop of interest.
We found that this feedback loop leads to bistability between weak and
strong adhesion and to strong spatial gradients in the deformation
pattern. In fact, the experimental results can be only explained by
incorporating the feedback loop. Our model shows how coupling of
mechanics and biochemistry can be modeled in general. In the future,
it might be used to address also other issues in the context of cell
adhesion, including cell adhesion to soft elastic substrates or to
cyclically stretched substrates.

In our model, spatial variations in the deformation pattern result
from inhomogeneous reaction-diffusion fields. This prediction should
be experimentally tested in the future. Indeed it has already been
reported that MLC-phosphorylation is larger at the periphery than at
the center (both before and after stimulation with calyculin)
\cite{c:pete04}, exactly as predicted by our model. Our model does not
account for inhomogeneities resulting from spatial variations in
mechanical properties (e.g.\ local accumulation of myosin II, actin or
$\alpha$-actinin), although in principle the model is capable to
describe these.  A detailed experimental analysis including detailed
measurements of local actin and myosin accumulation would be needed to
disentangle the relative contributions of biochemical and mechanical
factors to the inhomogeneous deformation pattern.

An intriguing aspect of our model is the way different stress
fibers might cooperate inside a living cell. Conceptually it is
easy to generalize our model to describe a system in which many
stress fibers share the signaling input and many focal adhesions
share the mechanical output. However, it remains a challenge to
model also the dynamics of the actomyosin system if it is not
completely condensed into stress fibers.

\textit{Acknowledgements} We thank Dominik Meidner for helpful
discussions on numerically solving partial differential equations.
This work was supported by the Center for Modeling and Simulation in
the Biosciences (BIOMS) at Heidelberg.

%\bibliography{ab}
%\bibliographystyle{unsrt}

\cleardoublepage
\begin{table}[ht]
\begin{center}
\renewcommand{\baselinestretch}{1}\normalsize
\small
\begin{tabular}{|l | p{5.2cm} | p{6cm}|}
\hline
\multicolumn{1}{|c|}{Abbreviation} & \multicolumn{1}{c|}{full name} &
\multicolumn{1}{c|}{function}\\
\hline
\hline
GDP/GTP & guanosine diphosphate/ \newline  guanosine triphosphate & small
molecule without and with a third phosphate group, energy source of conformational changes \\
\hline
GEF     & guanine nucleotide \newline exchange factor   & activates GTPases by
exchanging                                                                 GDP
for GTP \\
\hline
GAP     & GTPase-activating protein           & stimulates
GTP-hydrolysis, converting                                                       
   active GTPases (GTP-bound) to their inactive (GDP-bound) form\\
\hline
GDI     & guanine nucleotide \newline dissociation inhibitor & binds to inactive
form of GTPases,                                                         the
complex is soluble in the cytoplasm\\
\hline
MLC     & myosin light chain                            & subunit of myosin II molecular motors,
regulates myosin                                                          
binding to actin filaments\\
\hline
MLCK    & myosin light chain kinase                     & phosphorylates MLC\\
\hline
MLCP    & myosin light chain phosphatase                & dephosphorylates MLC
\\
\hline
MLCP-P  & phosphorylated MLCP                           & inactive form of MLCP \\
\hline
MBS     & myosin-binding subunit                      & subunit of MLCP whose phosphorylation
makes MLCP inactive \\
\hline
ROCK    & Rho-associated kinase                         & ROCK phosphorylates
MLCP at its                                                                
myosin-binding subunit (MBS)\\
\hline
I       & effect of calyculin & inhibits MLCP from dephosphorylating myosin, thus enhancing
contractility \\
\hline
\end{tabular}
\renewcommand{\baselinestretch}{1}\normalsize
\caption{Abbreviations and full names of the biochemical components. For each component,
a short description of its function is given.}
  \label{tab_comp}
\end{center}
\end{table}

\cleardoublepage
\begin{table}[ht]
\begin{center}
\renewcommand{\baselinestretch}{1.80}\normalsize
\begin{tabular}{|l c l c|}
\hline
\multicolumn{4}{|c|}{Model equations}\\
\hline
$ \frac{\partial ROCK(t)}{\partial t}$ & = & $\frac{r_1
F_{b}(t)(ROCK_{tot}-ROCK(t))}{K_1+(ROCK_{tot}-ROCK(t))} -\frac{V_{-1}
ROCK(t)}{K_{-1}+ROCK(t)}$
& (m1)\\
$\frac{\partial \textit{MLCP}(x,t)}{\partial t}$ & = & $\frac{V_{-2}
\textit{MLCP-P}(x,t)}{K_{-2}+\textit{MLCP-P}(x,t)}+D\frac{\partial^2
\textit{MLCP}(x,t)}{\partial x^2}$
& (m2)\\
$\frac{\partial \textit{MLCP-P}(x,t)}{\partial t}$ & = & $-\frac{V_{-2}
\textit{MLCP-P}(x,t)}{K_{-2}+\textit{MLCP-P}(x,t)}+D_{p}\frac{\partial^2
\textit{MLCP-P}(x,t)}{\partial x^2}$
& (m3)\\
$\frac{\partial n(x,t)}{\partial t}$ & = & $\frac{V_3
\left(1-n(x,t)\right)}{K_3+(1-n(x,t))} -\frac{r_{-3} \textit{MLCP}(x,t)  
n(x,t)}{K_{-3}I+n(x,t)}$
& (m4)\\
\multicolumn{3}{|l}{
$\left(\frac{\partial}{\partial x}\eta_{e}(x,t)\frac{\partial}{\partial x}
\frac{\partial}{\partial t}+\frac{\partial}{\partial x} k(x)
\frac{\partial}{\partial x}\right)u(x,t)=-\frac{1}{a}\frac{\partial}{\partial x}
F_{stall}(x,t)
$}
& (m5)\\
\hline
\multicolumn{4}{|c|}{Boundary conditions at x=0,L}\\
\hline
$\frac{\partial \textit{MLCP}(x,t)}{\partial x}$ & = & $ \pm \frac{V_2}{D}\frac{
ROCK(t) \textit{MLCP}(x,t)}{K_2+\textit{MLCP}(x,t)}$
& (bc2)\\
$\frac{\partial \textit{MLCP-P}(x,t)}{\partial x}$ & = & $ \mp
\frac{V_2}{D_{p}}\frac{ ROCK(t) \textit{MLCP}(x,t)}{K_2+\textit{MLCP}(x,t)}$
& (bc3)\\
\multicolumn{2}{|l}{$u(x,t)=0$} & \,\,stiff boundaries
& (bc5)\\
\hline
\multicolumn{4}{|c|}{Abbreviations}\\
\hline
\multicolumn{3}{|l}{$F_{stall}(x,t)=F_{max}\,n(x,t)$}
& \\
\multicolumn{3}{|l}{$\eta_{e}(x,t)=\eta(x)+F_{stall}(x,t)/v_0$}
& \\
\multicolumn{3}{|l}{$F_{b}(t)= a\eta_{e}(0,t)\partial_x \partial_t u(0,t)+ak(0)
\partial_x u(0,t)+F_{stall}(0,t)$}
& \\
\hline
\end{tabular}
\renewcommand{\baselinestretch}{1}\normalsize
\caption{Summary of model equations. Eqs.~(m1-m4) describe successive
  biochemical signaling events: (m1) focal adhesion associated
  activation of ROCK; (m2) and (m3) phosphorylation and diffusion of
  MLCP and dephosphorylation and diffusion of MLCP-P; (m4) regulation
  of the active fraction of the myosins, which is identified with the
  phosphorylated fraction of MLC. Eq.~(m5) is the mechanical model
  equation for stress fibers, where $u(x,t)$ is the displacement along
  the fiber.  The boundary conditions for the partial differential
  Eqs.~(m2), (m3), (m5) are given by (bc2), (bc3), (bc5) respectively.
  In Eq.~(bc2), (bc3), the upper (lower) sign is valid for the left
  $x=0$ (right $x=L$) boundary. For the sake of clarity, we have
  introduced the listed abbreviations for the stall force $F_{stall}$,
  the effective viscosity $\eta_{e}$ and the force exerted onto the
  boundary $F_{b}$. The presented results have been derived with the
  assumptions that: (I) The diffusion properties of the phosphorylated
  and unphosphorylated form of the phosphatase are the same, hence
  $D=D_{p}$. (II) The viscoelastic properties of the stress fiber do
  not vary in space, therefore $k(x)\rightarrow k$ and $\eta_{e}(x,t)
  \rightarrow \eta+F_{max}\,n(x,t)/v_0$.}
  \label{tab_model_eq}
\end{center}
\end{table}

\cleardoublepage
\begin{table}[ht]
\renewcommand{\baselinestretch}{0.50}\normalsize
\tiny
%\scriptsize
%\footnotesize
\begin{tabular}{|l|l|r|r|r|}
\hline
\multicolumn{5}{|c|}{Time dependent reaction variables}\\
\hline
\multicolumn{1}{|c|}{Abbreviation} & \multicolumn{1}{c|}{Meaning} &
\multicolumn{1}{c|}{Used value} & \multicolumn{1}{c|}{Reference values} &
\multicolumn{1}{c|}{References}\\
\hline
$ROCK$             & activated form of ROCK                        &  
$0 \dots 5nM$       &$\gtrsim1nM$               &\cite{Feng99}      \\
$\textit{MLCP}$              & unphosphorylated form of MLCP
                  &   $0 \dots 1.2\mu M$  &$1.2\pm0.3\mu M$&\cite{Hartshorne98}
\\
\textit{MLCP-P}    & phosphorylated form of MLCP                    &   $0 \dots
1.2\mu M$  &$1.2\pm0.3\mu M$&\cite{Hartshorne98} \\
$n$               & fraction of active myosin         &   $0 \dots 1$        
&[MLC-phos]/[myosin] &   \\

\hline
\hline
\multicolumn{5}{|c|}{Reaction constants}\\
\hline
$MLCK$            & myosin light chain kinase             &   $0.1\mu M$        
 &$\gtrsim 100 nM$           &\cite{Nagamoto84}  \\
$M$               & myosin concentration                  &   $    30\mu M$     
 &$ 25\dots30\mu M$          &\cite{Butler94}    \\
$K_1$           & Michaelis constant                    &   $5nM$              
& \textbf{(no value)}       &                   \\
$K_{-1}$        & Michaelis constant                    &   $4.7nM$            
& \textbf{(no value)}       &                   \\
$K_2$           & Michaelis constant                    &   $0.1\mu M$         
&$0.10\pm0.01\mu M$         &\cite{Feng99}      \\
$K_{-2}$        & Michaelis constant                    &   $15\mu M$          
& \textbf{(no value)}       &                   \\
$K_3*$M         & Michaelis constant                    &   $20\mu M$          
&$52.1\pm7.1\mu M$          &\cite{Amano96}     \\
                &                                       &                      
&$34.5\pm2.8\mu M$          &\cite{Feng99}      \\
                &                                       &                      
&$18\mu M$                  &\cite{Hathaway79}  \\
                &                                       &                      
&$7.7 \dots 96.0\mu M$      &\cite{Nunnally85}  \\
                &                                       &                      
&$19 \dots 53\mu M$         &\cite{Nagamoto84}  \\
                &                                       &                      
&$20\mu M$                  &\cite{Bartelt87}      \\
$K_{-3}*$M      & Michaelis constant                    &   $10\mu M$          
&$10 \mu M$                 &\cite{Pato83}      \\
$r_{1}$         & rate constant                         &   $0.3nM/s\,nN$      
& \textbf{(no value)}       &                   \\
$V_{-1}$        & maximum velocity                      &   $1.8nM/s$          
& \textbf{(no value)}       &                   \\
$r_2$           & rate constant                         &   $2.4\ 1/s$   
&$2.36\pm0.10\,\ 1/s$ &\cite{Feng99}      \\
$R_{2}$         & maximum velocity                      &$4.8\mu
m/s$& $r_2*v_{b}$             &                   \\
$V_{-2}$        & maximum velocity                      &   $0.1\mu M/s$       
& \textbf{(no value)}       &                   \\
$r_3*$M         & rate constant                         &   $10\ 1/s$    
&$2.00\pm0.36\,\ 1/s$ &\cite{Amano96}     \\
                &                                       &                      
&$3.85\pm0.095\,\ 1/s$&\cite{Feng99}      \\
                &                                       &                      
&$5.17\,\ 1/s$        &\cite{Hathaway79}  \\
                &                                       &                      
&$7.37\dots171.3\,\ 1/s$&\cite{Nunnally85}\\
                &                                       &                      
&$70\dots100\,\ 1/s$  &\cite{Nagamoto84}  \\
                &                                       &                      
&$4.64\,\ 1/s$        &\cite{Bartelt87}   \\
$V_3$           & maximum velocity                      &   $1.0\mu M/s$       
&$r_3*$MLCK                 &                   \\
$r_{-3}*$M      & rate constant                         &   $21\ 1/s$    
&$21\ 1/s$            &\cite{Pato83}      \\
$D$             & diffusion constant of \textit{MLCP} \& \textit{MLCP-P}       &
  $14\mu m^2/s$       &$10 \dots 100\mu m^2/s$    &\cite{Kenworthy01} \\
$v_{b}$         & effect. react. vol. of FAs            &   $2.0 \mu m$        
& \textbf{(no value)}       &                   \\
$I$             & inhibition parameter                  &   $1\rightarrow 3$   
&\textbf{(no value)}        &                   \\
\hline
\hline
\multicolumn{5}{|c|}{Parameters of mechanical model}\\
\hline
$F_{max}$       & stall force                           &   $50nN$             
&\textbf{(no value)}        &                   \\
$v_0$           & maximum motor velocity                &   $1.0 \mu m/s$      
&$\approx0.1\dots1 \mu m/s$ &\cite{Sellers90}   \\
a               & vec-element spacing                   &   $1.0 \mu m$        
&$1.0 \mu m$                &\cite{c:pete04}  \\
k               & spring stiffness                      &   $45 nN/\mu m$      
&$45.7 nN/a$                &\cite{Deguchi06}   \\
$\eta$          & viscosity                             &   $45 nN\,s/\mu m $  
&$\approx\tau k=45.7nN\,s/a$&\cite{Deguchi06,Kumar06}\\
L               & fiber length                          &   $50 \mu m$         
&$\approx20\dots80\mu m$    &                   \\
\hline
\end{tabular}
\renewcommand{\baselinestretch}{1}\normalsize
\caption{Model parameters based on literature search. We have set the model
  parameters such that they fit into the reported range.
  The equation for the phosphorylated
  fraction of MLC is normalized to the total myosin concentration
  denoted by $M$. In order to make the involved reaction constant
  comparable to the literature values we give $K_3$, $K_{-3}$, $r_3$
  and $r_{-3}$ scaled with $M$. \eq{eq_ROCK} translates
  mechanical forces into biochemical activation. For this reason the
  units of the rate constant $r_1$ are given as
  $[nM/s\,nN]$. The typical relaxation time, $\tau$, of stress fibers
  is of the order of a few seconds \cite{Kumar06} therewith
  we roughly estimate the viscosity value as $\eta\approx\tau
  k$ where we use $\tau=1s$. Myosin activation by
  calyculin is modelled as competitive inhibition of the phosphatase.
  The inhibition parameter $I$ is switched instantaneously from
  $I=1$ to $I=3$. Some of the  reported values have been
  measured for the interactions of protein fractions and not for the
  native proteins. Furthermore, the experiments have been done on
  proteins extracted from different species.}
        \label{tab_parameters}
\end{table}

\cleardoublepage
%-------------------------------------------------------------

\begin{figure}[h] \renewcommand{\baselinestretch}{1}
        \centerline{\epsfysize 10cm \epsfbox{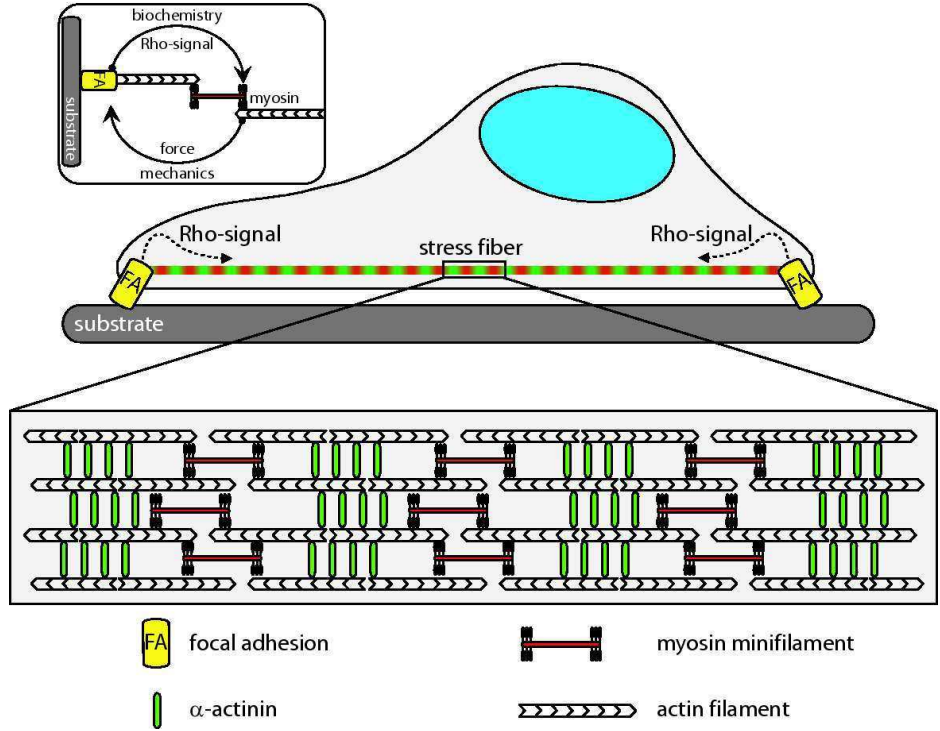}}
        \caption{Cells adhere to the extracellular matrix by
          integrin-mediated contacts called focal adhesions. These
          contacts are the anchor points of stress fibers, which are
          actin filament bundles held together by the crosslinker
          molecule $\alpha$-actinin and myosin II molecular motors.
          The myosins are assembled in myosin mini-filaments. Due to
          myosin motor activity stress fibers are under tension and
          exert forces to focal adhesions.  This mechanical stimulus
          initiates biochemical signals (Rho-signal) that originate
          from focal adhesions and propagate into the cytoplasm,
          altering in turn myosin activity.  Therefore the system of
          focal adhesions and stress fibers are connected by a
          biochemical and mechanical positive feedback loop (inset).
          The spatial part of our model is one-dimensional with one
          stress fibers extending between the two focal adhesions at
          $x=0$ and $x=L$.}
        \label{fig_overview}
\end{figure}
%-------------------------------------------------------------

%-------------------------------------------------------------
\begin{figure} \renewcommand{\baselinestretch}{1}
        \centerline{\epsfysize 10cm \epsfbox{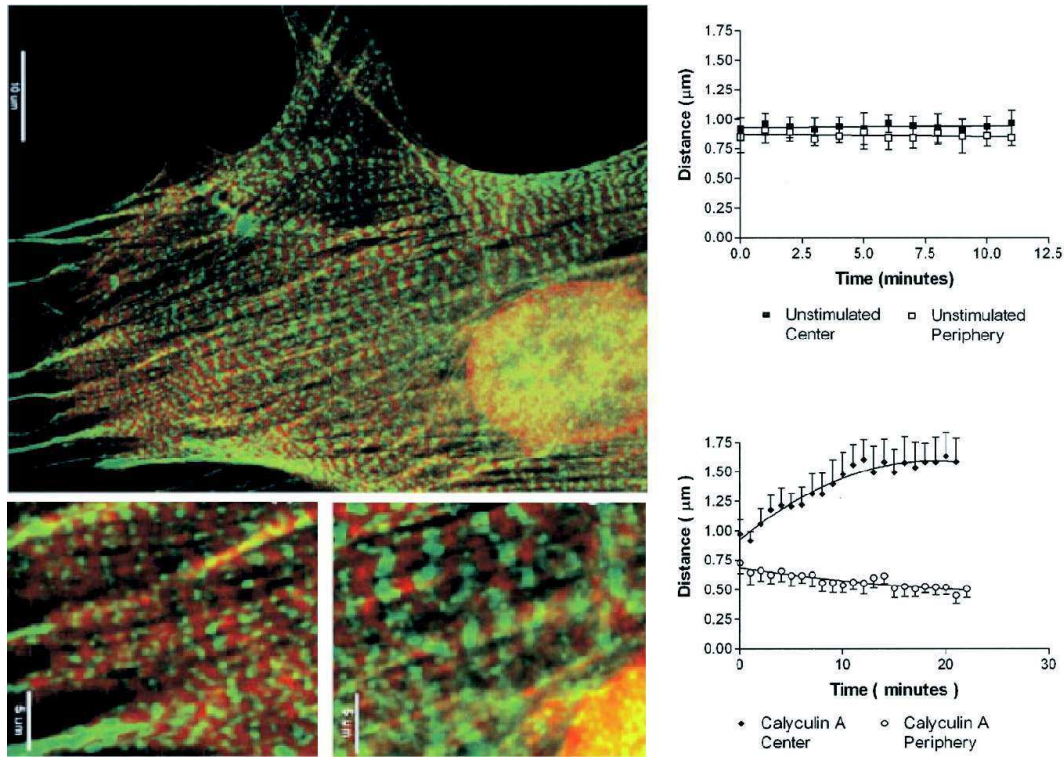}}
        \caption{(A) Peterson et al. \cite{c:pete04} studied the
          deformation of stress fibers in fibroblasts by using
          fluorescently labelled $\alpha$-actinin (in green) and
          myosin light chain (in red). These two components arrange
          sequentially along the stress fibers and thereby form
          regular striation patterns.  Myosin contractility was
          stimulated with the drug calyculin A. Then an inhomogeneous
          striation pattern results: stress fibers contract at the
          cell periphery (B) but expand at the cell center (C).  These
          results were quantified by time course measurements of the
          mean pattern bandwidths in the respective regions. Compared
          with the control (D), stimulation of contractility leads to
          very strong spatial gradients (E) on the time scale of tens
          of minutes.}% end caption
        \label{fig_exp_data}
\end{figure}
%-------------------------------------------------------------

%-------------------------------------------------------------
\begin{figure} \renewcommand{\baselinestretch}{1}
        \centerline{\epsfysize 10cm \epsfbox{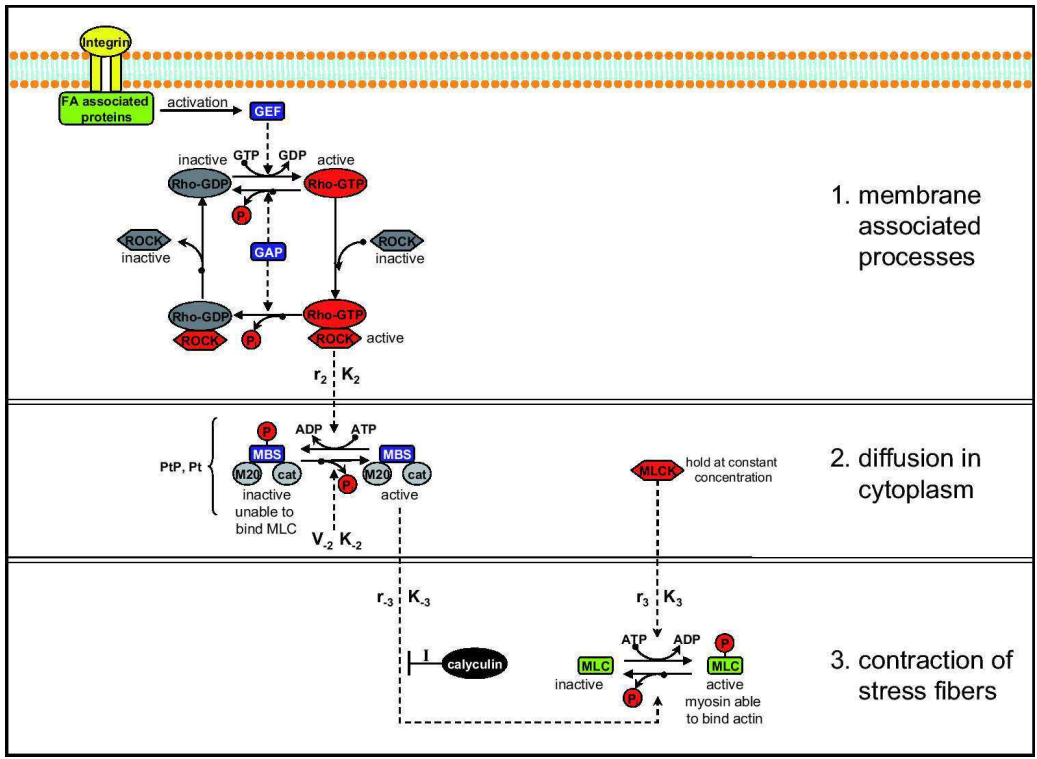}}
        \caption{Signaling pathway that controls myosin contractility
          depicted in its appropriate spatial context: mechanical cues
          are transduced to various biochemical signals at focal
          adhesions, however the precise mechanisms have not been
          resolved yet. One possible mechanism is that a Rho-GEF is
          activated by a mechanosensitive process at focal adhesions.
          Rho-GEF then promotes Rho-GTP loading and subsequent
          complexation with Rho-associated kinase (ROCK) which gets
          activated. Active ROCK is able to phosphorylate myosin light
          chain phosphatase (MLCP) at its myosin-binding subunit
          (MBS). MLCP and MLCP-P are freely diffusible in the
          cytoplasm and thus can reach the myosins in the stress
          fibers.  Increased phosphorylation of MLCP to MLCP-P by ROCK
          effectively leads to increased phosphorylation of myosin
          light chain (MLC), thus increasing myosin
          contractility.}% end caption
        \label{fig_biochemistry}
\end{figure}
%-------------------------------------------------------------

%-------------------------------------------------------------
\begin{figure} \renewcommand{\baselinestretch}{1}
        \centerline{\epsfysize 7cm \epsfbox{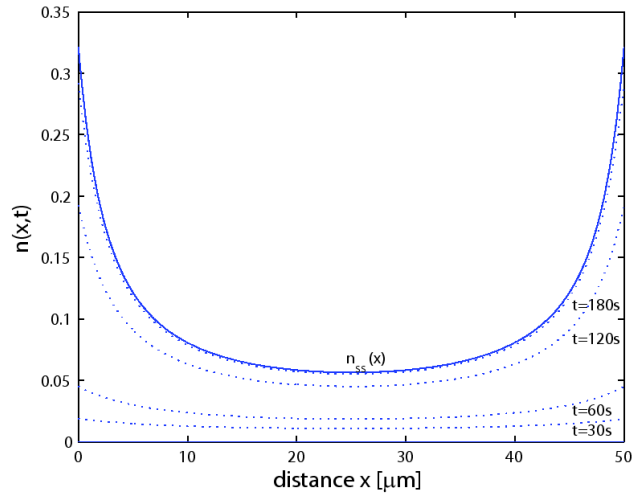}}
        \caption{Spatial dependence of the active myosin fraction
        $n(x,t)$ at four different time points $t\in\{30s, 60s, 120s, 180s\}$
        as well as for the steady state, $n_{ss}(x)$:
        We solve the biochemical model implying the
        artificial initial conditions that all components are at
        zero activation level, but set the boundary force to $5nN$.
        Because of this mechanical trigger at focal adhesions, MLC gets
        preferentially activated at the boundaries via the
        Rho-pathway which leads to a steady increase of the
        myosin activation level. Due to diffusible compounds in the
        Rho-pathway, the increased activation level is smoothed out
        towards the center of the cell.
        }% end caption
        \label{fig_n_para_plot}
\end{figure}
%-------------------------------------------------------------

%-------------------------------------------------------------
\begin{figure} \renewcommand{\baselinestretch}{1}
        \centerline{\epsfysize 7cm \epsfbox{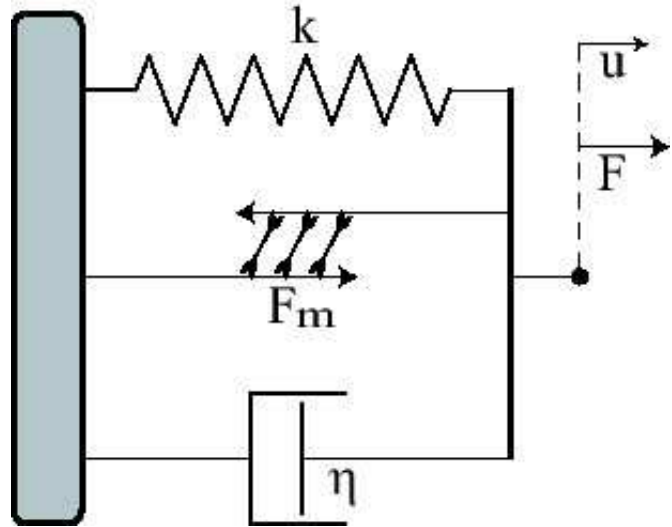}}
        \caption{The viscoelastic and contractile element
          (vec-element) consists of a spring of stiffness $k$, a
          contractile module of contraction force $F_m$ and a dashpot
          of viscosity $\eta$ that are all connected in parallel.
          Spring and dashpot taken alone are the usual Kelvin-Voigt
          model of viscoelastic material. The properties of the
          contractile module are characterized to first order by the
          linearized force-velocity relation.}
        \label{fig_vec-element}
\end{figure}
%-------------------------------------------------------------

%-------------------------------------------------------------
\begin{figure} \renewcommand{\baselinestretch}{1}
        %\centerline{\epsfysize 7cm\epsfbox{fig_springs_dashpots_and_motors_old.eps}}
        \centerline{\epsfysize 11.5cm
\epsfbox{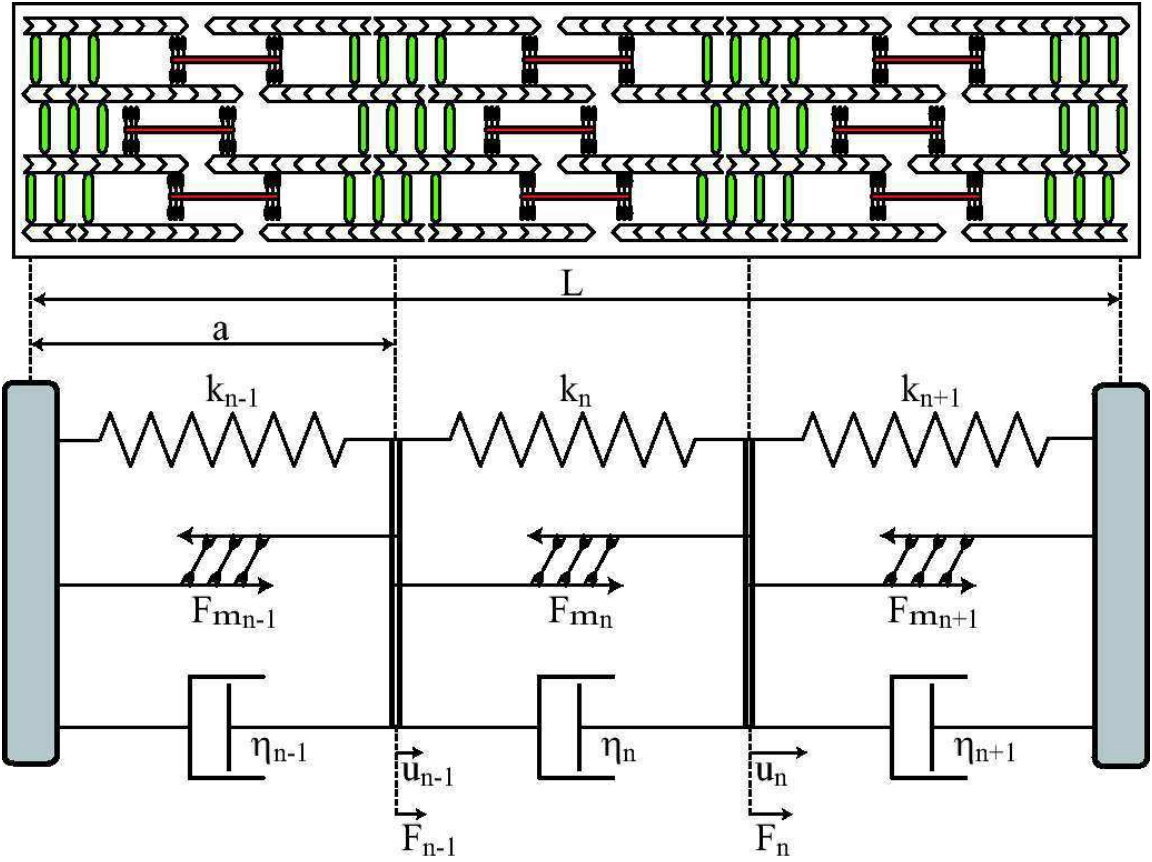}}
        \caption{We model a stress fiber as a string of
        vec-elements such that the spring stiffness $k_n$,
        the viscosity $\eta_n$ and the motor force $F_{m_n}$ can
        vary spatially. E.g. the latter will vary spatially due to
        different myosin activity at the periphery compared to the
        center. The displacement of a certain site $n$ is denoted by
        $u_n$. The force $F_n$ onto this site $n$ consists of
        elastic, viscous and motor contributions, compare
        \eq{eq_Fn}. In the following we assume fix boundary
        conditions, namely $u(0,t)\equiv0$ and $u(L,t)\equiv0$ where
        $L$ is the total length of the fiber.
        }% end caption
        \label{fig_springs_dashpots_and_motors}
\end{figure}
%-------------------------------------------------------------

%-------------------------------------------------------------
\begin{figure} \renewcommand{\baselinestretch}{1}
        \centerline{\epsfysize 7cm \epsfbox{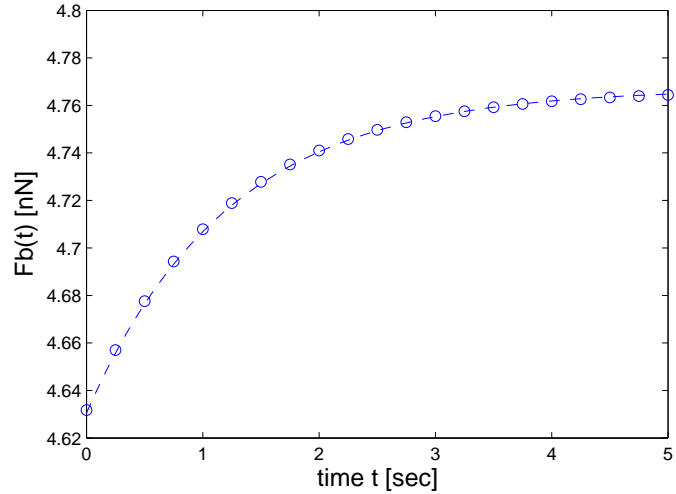}}
        \caption{Time course of boundary force $F_{b}(t)$. The
          solution of the Volterra equation, \eq{eq_volterra}, indicated
          by circles, was calculated by applying an iteration rule
          further explained in the main text. The dashed line is
          the boundary force deduced from direct numerical
          solution of \eq{eq_boxed} which we include for comparison.
          For the assumed initial condition, $\partial_{x}u(x,0) = 0$,
          the boundary force increases from its non-zero value at $t=0$, given by
          \eq{eq_Flb0} and quickly saturates at somehow larger values.}
        \label{fig_fb_ana_num}
\end{figure}
%-------------------------------------------------------------

%-------------------------------------------------------------
\begin{figure} \renewcommand{\baselinestretch}{1}
        \centerline{\epsfysize 7cm \epsfbox{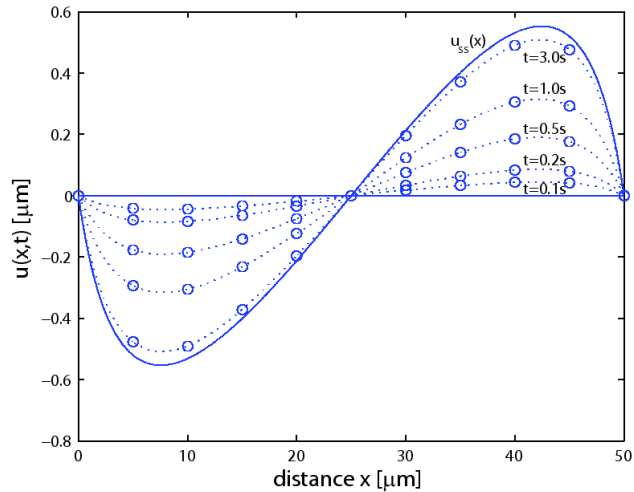}}
        \caption{We first analyze the solution of the mechanical
          equation \eq{eq_boxed} by assuming a steady state myosin
          activation level, $n_{ss}(x)$, shown in
          \fig{fig_n_para_plot}. For this simplifying case where the
          myosin activation level is not time dependent, \eq{eq_boxed}
          can be solved both numerically and analytically.
          \fig{fig_u_para_plot} shows the analytical solution,
          \eq{eq_ana_sol}, indicated by circles whereas the direct
          numerical solution of \eq{eq_boxed} is indicated by dotted
          lines.  The solution is given at time points
          $t\in\{0.1s,0.2s,0.5s,1.0s,3.0s\}$ as well as for the steady
          state $u_{ss}(x)$, assuming the initial condition:
          $\partial_{x}u(x,0)\equiv0$ and the boundary conditions:
          $u(0,t)\equiv0$ and $u(L,t)\equiv0$.}
        \label{fig_u_para_plot}
\end{figure}
%-------------------------------------------------------------

%-------------------------------------------------------------
\begin{figure} \renewcommand{\baselinestretch}{1}
        \centerline{\epsfysize 7cm \epsfbox{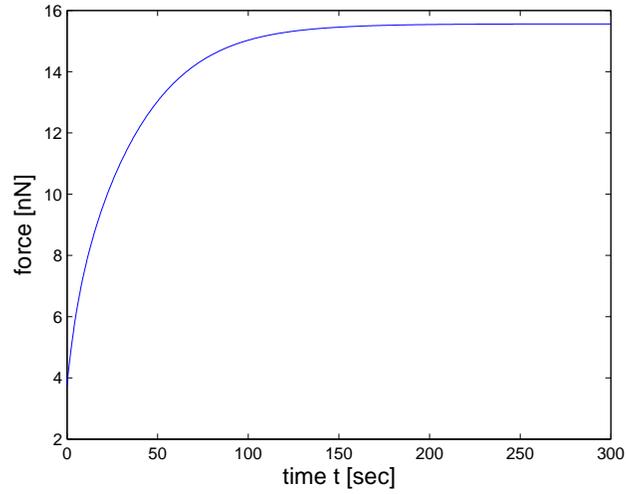}}
        \caption{At $t=0$ calyculin is added to the system.
          Then myosin motors along the filament get
          activated and further increase the tension within the fiber. The
          time course of the force $F_b(t)$ transduced to the
          boundaries is shown.}
        \label{fig_fb_inhib}
\end{figure}
%-------------------------------------------------------------

%-------------------------------------------------------------
\begin{figure} \renewcommand{\baselinestretch}{1}
        \centerline{\epsfysize 7cm \epsfbox{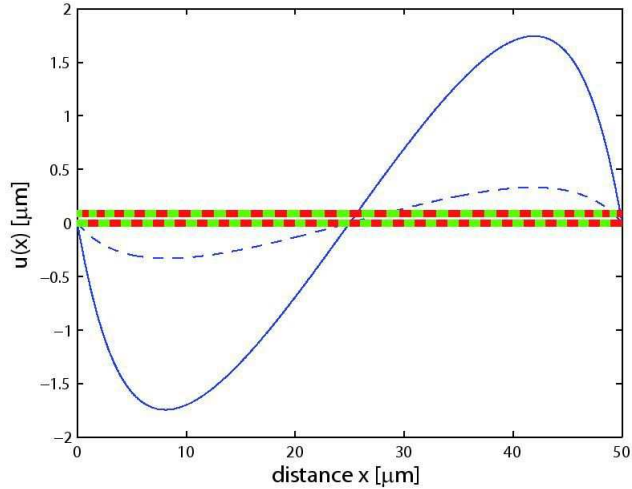}}
        \caption{Steady state solution of the displacement $u(x)$
          along the fiber before (dashed line) and after (solid line)
          stimulation with calyculin. The stimulation strongly
          increases the deformation of the fiber resulting in
          substantial distortion of the expected striation pattern
          (upper line) compared to the striation pattern of a
          completely undistorted fiber (lower line). The upper
          striation pattern was calculated from the displacement data.
          The bands close to the boundaries have been contracted
          (about 55\%) whereas the bands around the center have been
          expanded (about 15\%), compare also
          \fig{fig_density_inc_wo}.}
        \label{fig_u_b_n_a}
\end{figure}
%-------------------------------------------------------------

%-------------------------------------------------------------
\begin{figure} \renewcommand{\baselinestretch}{1}
        \centerline{\epsfysize 7cm \epsfbox{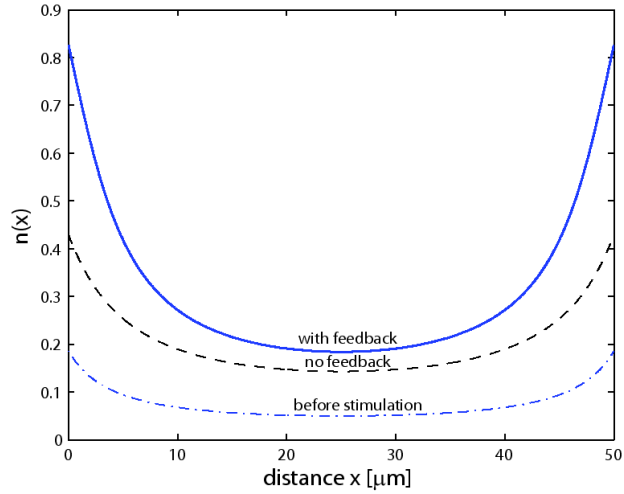}}
        \caption{Steady state profile of the
        active myosin fraction before stimulation with the drug
        (dash-dotted line), after stimulation including
        the mechanical feedback (solid line) and after stimulation
        but neglecting the mechanical feedback (dashed line).
        For the latter case, homogeneous induction of the drug
        cause an almost uniform elevation of myosin activity.
        Slight differences between center and periphery persist
        but rather marginally extend, whereas the closed feedback
        system result in an amplification of the spatial
        differences of myosin activation.
        }% end caption
        \label{fig_nss_inc_wo}
\end{figure}
%-------------------------------------------------------------

%-------------------------------------------------------------
\begin{figure} \renewcommand{\baselinestretch}{1}
        \centerline{\epsfysize 7cm \epsfbox{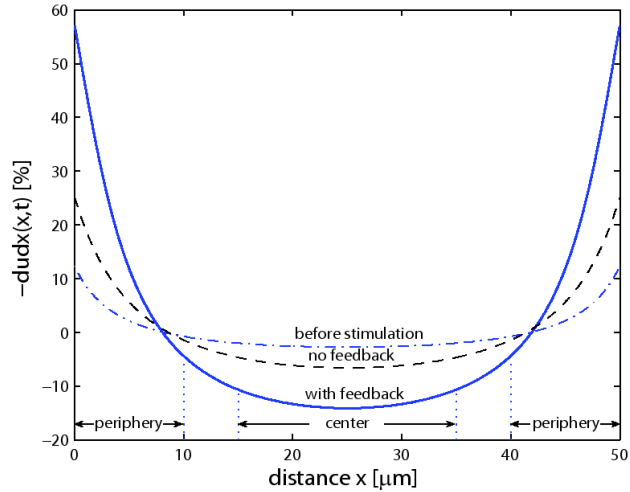}}
        \caption{Relative change in density along
          the fiber in the steady state before stimulation
          with the drug (dashed-dotted line), after stimulation
          with the drug including the feedback (solid line)
          and after stimulation but neglecting the mechanical
          feedback (dashed line). Positive values correspond to
          a compression of the fiber, whereas negative values indicate
          elongation. In case of the closed feedback (solid line)
          the fiber strongly contracts close to the boundaries
          up to 55\% but elongates at the center to about 15\%.
          In order to compare the model results with the experimental
          findings, we arbitrarily define central (center $\pm10\mu m$)
          and peripheral (edges $\pm10\mu m$) regions of the cell,
          indicated by vertical lines.
          }% end caption
        \label{fig_density_inc_wo}
\end{figure}
%-------------------------------------------------------------

%-------------------------------------------------------------
\begin{figure} \renewcommand{\baselinestretch}{1}
        \centerline{\epsfysize 7cm \epsfbox{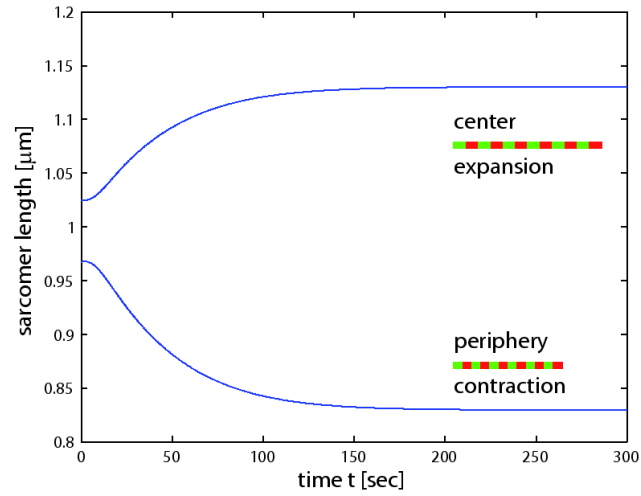}}
        \caption{Time courses of the mean bandwidth of the fiber in the center
(upper
          curve) and in the periphery of the cell (lower curve): The
          shown values are averages of the bandwidths at the
          center $\pm10\mu m$ for the central region and at the edges
          $\pm10\mu m$ for the peripheral region. The defined
          intervals are also shown in
          \fig{fig_density_inc_wo}. The expected steady
          state striation patterns for the two distinct regions are
          shown as insets. These results agree qualitatively with
          the experimental findings by Peterson et al., shown in
          \fig{fig_exp_data}.
          }% end caption
        \label{fig_mean_bw}
\end{figure}
%-------------------------------------------------------------
\end{document}